\documentclass[aps,twocolumn,superscriptaddress,10pt,fleqn,floatfix]{revtex4}
\usepackage{graphicx}
\usepackage{dcolumn}
\usepackage{bm}

\begin{document}
\author{A.Avdeenkov}
\affiliation{Institut f\"ur Kernphysik, Forschungszentrum J\"ulich, 52425 J\"ulich, Germany}
\affiliation{Institute of Physics and Power Engineering, 249020 Obninsk, Russia}
\author{F.Gr\"ummer}
\affiliation{Institut f\"ur Kernphysik, Forschungszentrum J\"ulich, 52425 J\"ulich, Germany}
\author{S.Kamerdzhiev}
\affiliation{Institute of Physics and Power Engineering, 249020 Obninsk, Russia}
\author{S.Krewald}
\affiliation{Institut f\"ur Kernphysik, Forschungszentrum J\"ulich, 52425 J\"ulich, Germany}
\author{N.Lyutorovich}
\affiliation{Institut f\"ur Kernphysik, Forschungszentrum J\"ulich, 52425 J\"ulich, Germany}
\affiliation{Institute of Physics S.Petersburg University, Russia}
\author{J.Speth}
\affiliation{Institut f\"ur Kernphysik, Forschungszentrum J\"ulich, 52425 J\"ulich, Germany}
\affiliation{Institute of Nuclear Physics, PAN, PL-31-342 Cracow, Poland}

\title{Self-consistent calculations within the Extended Theory of Finite Fermi Systems }


\begin{abstract}
The \emph{Extended Theory of Finite Fermi Systems}(ETFFS) describes nuclear excitations considering
phonons and pairing degrees of freedom, using the effective Landau-Migdal interaction and nuclear 
mean fields obtained from experimental data. Here we employ the nuclear mean field derived from Skyrme 
interactions and the corresponding particle hole interaction.
This allows to extend the range of applicability of the ETFFS to experimentally not yet investigated short-lived isotopes.
We find that Skyrme interactions which reproduce at the mean field level both ground state properties 
and nuclear excitations  are able to describe the spreading widths
of the giant resonances in the new approach, but produce shifts of the centroid energies.
A renormalization of the Skyrme interactions is required for approaches going beyond the
mean field level. 
\end{abstract}
\maketitle

\section{introduction}
The renewed interest in nuclear structure physics is due to its application to astrophysics.
Theories for the element synthesis in stars need both bulk properties of unstable isotopes like
masses, and more specific information, such as level densities and neutrino-nucleus cross
sections.
Eventually, the compression modulus and the symmetry energy of neutron-rich nuclei will be
determined. For a recent review, see e.g. Ref. \cite{Thielemann:2001rn}.

Most calculations in this field are restricted to the mean field approximation
which describes excited states by the one-particle one-hole random phase approximation (1p1h RPA).
In this approximation one is able to describe two integral characteristics of giant resonances:
the mean energy and the integrated strength.
The width of the giant resonances is underestimated in mean field approaches, however,
even if one includes the single particle continuum.

In unstable nuclei, the analysis of, for example, the breathing mode and the giant electric dipole resonance
requires realistic strength distributions because these resonances are expected to be broad and
various multipole resonances overlap one another. Theories beyond the mean field level are called for. 

The \emph{Extended Theory of Finite Fermi Systems} (ETFFS) is a linear response theory which
incorporates  phonons and pairing degrees of freedom. 
One essentially replaces the particle-hole propagator, which leads to the random phase approximation, by a
more general propagator and changes, correspondingly, the renormalized interaction, the effective
operators and the ground-state correlations.
Details can be found in Refs. \cite{Tsel89,Kam91,kst93,kstw93}, where the authors
worked out the coupling of phonons to the single particle propagators and to the interaction in a consistent way.
The present calculations are based on the
\emph{Quasi particle Time Blocking Approximation} (QTBA) that also considers pairing correlations \cite{Vic}.
Examples and references to previous calculations and microscopic analyses of experimental data can
be found in \cite{rev04,gruemmer}. Another model, the QRPA plus phonon coupling model, has been developed by Bortignon and coworkers \cite{Bort01}.

Most  excited states predicted by the conventional random-phase approximation  depend  sensitively
on the level density of the single-particle spectrum, as is well known from the schematic model by Brown and Bolsterli \cite{BB59}.
As the TFFS, with and without pairing, uses the experimental single particle spectrum as far as it
is available, e.g. \cite{Ring74,Zawischa}, it generally comes closer to the experimental data than
self-consistent approaches do. 
For that reason an extended version of the TFFS in which the effects of the phonons are included
is the appropriate tool for systematic investigations of both  collective and non collective states \cite{Speth07}.
Its applicability, however, is limited by the need for experimental information on the level
density in the neighboring odd mass nuclei.

At present, these properties of most short-lived neutron rich nuclei are not known experimentally.
Therefore one is forced to use  nuclear mean fields extrapolated from the known stable nuclei. One of the most successful
approaches starts with Skyrme forces, a class of {\it effective interactions} defined at the
{\it mean field level}. When going beyond the mean field level, one has to expect a renormalization
of the interaction strengths \cite{Epelbaum:2005pn}.

In the present communication, we investigate quantitatively to which extent a renormalization
of the Skyrme parameterization may be required
by employing a self-consistent mean field and the corresponding particle hole interaction
derived from existing Skyrme forces in the framework of the Extended Theory of Finite Fermi systems.

There are various Skyrme parameterizations that reproduce the ground state properties of the known isotopes with
similar accuracy but result in different mean fields, as reviewed recently in
Refs. \cite{Bender:2003jk,Bertsch:2004us}.

The nuclear matter properties of the Skyrme parameterizations, such as the effective mass, the
compression modulus, and the symmetry energy  can be used to group the effective interactions into
some broad classes. The versions initiated by Vautherin and Brink were fitted to ground state
properties only. Those interactions are characterized by a large incompressibility (see SIII
in Table \ref{tab:t1}).
\begin{table}[!hb]
\caption{\label{tab:t1} Nuclear matter properties of the
Skyrme parameterizations used in the present calculation;
for comparison also SIII is given}
\begin{center}
\begin{tabular}{|c|c|c|c|c|}
\hline
  &  $m^{*}/m$   & $\kappa(MeV)$  & $a_s[MeV]$ \\
\hline
BSk5 &  0.92 & 237 & 28.7\\
SLy4 &  0.70 & 230 &32.0\\
SKM* &  0.79 & 218  &30\\
SIII &  0.76 & 356  &28.2\\
\hline
\end{tabular}
\end{center}
\end{table}
Later versions of the Skyrme interactions modified the density dependence
of the interaction in order to allow a smaller incompressibility. Those interactions
reproduce the mean excitation energies of giant resonances reasonably well and are 
therefore a natural starting point for our investigation.

The interaction SLy4 \cite{SLy4} has been used in(Q)RPA \cite{terasaki1,terasaki2,goriely1} and in some first
attempts to include explicitly long range quadrupole correlations {(\it alias quadrupole phonons)}
into the nuclear ground state \cite{Bender:2005ri}.
An investigation similar to the present one, but based on the relativistic mean field approach,
has been published recently \cite{LR06,LRT06}.
Their results, calculated with and without phonons differ very little from each other, in contrast
to our results. The reason is the so-called subtraction procedure of Ref. \cite{Vic} which these authors apply. 
This is an {\it ad hoc} condition that is difficult to justify, as one needs a consistent description of both 
even even and even odd nuclei.

\section{Method}
We calculate the mean field within the HFB approach using the code HFBRAD \cite{bennaceur} and
the excited states with the the QTBA version \cite{Vic} of the ETFFS.
Here, as mentioned before, the authors developed a renormalization procedure that corrects for possible double
counting of the phonon contributions that are already included in the Skyrme parameterization.
We leave this correction out.  This is also in line with the investigations by Bertsch et al. \cite{Bender:2005ri},
who argue that  an effective interaction has to be refitted if one includes additional correlations.
As usual, the residual interaction for (Q)RPA and the present theory is derived as the second
derivative of the Skyrme energy functional \cite{speth,KKSF77}.
In the present calculation we make some simplifications.
For the non magic nuclei we solved the HFB equations and extracted the quasi particle energies,
wave functions and the occupation numbers. 
In the second derivative, the ph-interaction, we dropped the spin-orbit term and we approximated
the velocity-dependent terms by its Landau-Migdal limit \cite{speth,KKSF77}.
There are two kinds of velocity-dependent terms: the first is $\propto \bf{k^2} \delta (r-r')$ and
the second one is $\propto \bf{k^\dag} \delta(r-r') \bf{k}$
(p-wave interaction in momentum space).
The value of the first term averaged  over the angular variables  gives
$\frac{1}{2}k_F^2\delta(r-r')$, while that of the second one vanishes.
Such an approximation, of course, violates self-consistency and one has to modify the parameters
of the residual interaction to obtain the spurious center-of-mass state at zero energy.
For this reason we changed the parameters proportional to $t_1 k_F^2 \delta (r-r')$
up to 20\%.

In principle, the ETFFS treats the single-particle continuum on the RPA level for magic nuclei
exactly and includes the ground state correlations caused by CC (GSC$_{phon}$) \cite{rev04}
explicitly.
However, because of technical difficulties caused by the pairing correlations, we did not consider
these effects in the present calculation. The ground state correlations are therefore restricted
to the conventional RPA correlations.
We discretized the continuum with a quasi-particle energy cutoff of 100 MeV.
We did not find any noticeable differences in the results with a larger basis.
We used 14-16 low-lying phonons with angular momentum $L=2-6$ and natural parity.

\section{Results}
\subsection{Collective isovector electric dipole states}
Here we present our results for the giant dipole excitation for some of the tin isotopes from
$^{100}$Sn up to $^{132}$Sn, where we used the SLy4 \cite{SLy4} parameterization of the
Skyrme force. The SLy4 \cite{SLy4} parameterization is not only adjusted to ground state properties, but it also
reproduces within the RPA the GDR and isoscalar collective states reasonably well. For comparison we used for $^{132}$Sn two further parameterizations, namely BSk5 \cite{BSk5} and SKM* \cite{SkM*}. We compare our theoretical results with the data of $^{120}$Sn and $^{132}$Sn. In general, the experimental GDR data are analysed by fitting to a Lorentzian.
For that reason we have used the same procedure to determine the calculated integral
characteristics \cite{kst93,tselyaev00}.
The mean energies E$_{0}$ and the resonance widths ${\Gamma}$ have been extracted from the calculated strength distribution under the condition
that the first three energy-weighted moments of the Lorentzian and the theoretical distribution of
the total photo absorption cross section should coincide.
This method works well in the case of the ETFFS results.

\begin{figure}
{\includegraphics[width=.95\linewidth,angle=-0]{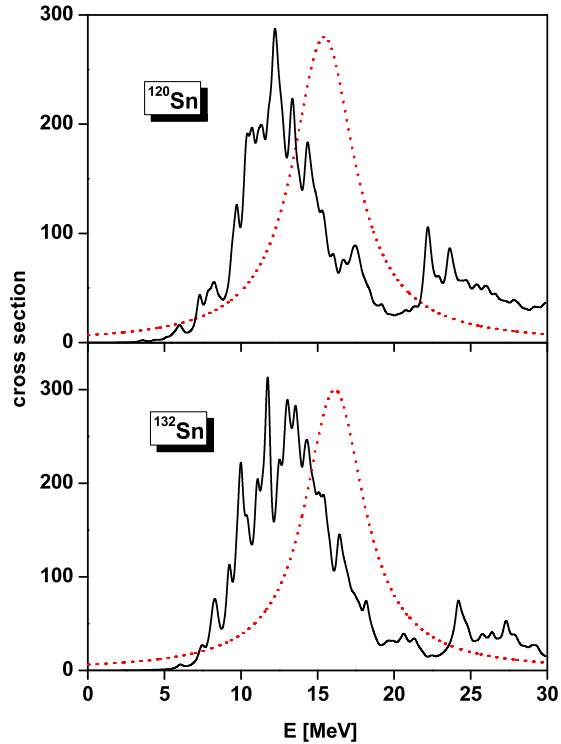}}
\caption{\label{fig1}E1 photo absorption cross section for $^{120}$Sn and $^{132}$Sn calculated
within the present theory (full). The dotted lines indicate the data for $^{120}$Sn \cite{Fultz69} and $^{132}$Sn \cite{GSI}.
The smearing parameter is $\Delta =0.2$ MeV. The SLy4 Skyrme parametrization was used. }
\end{figure}

\begin{figure}
\centerline{\includegraphics[width=.95\linewidth,angle=-0]{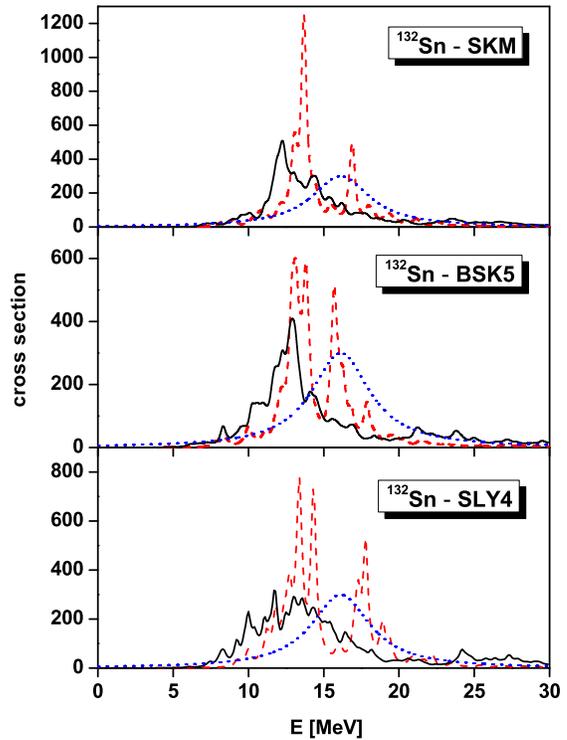}}
\caption{\label{fig2}E1 photo absorption cross section for $^{132}$Sn calculated within the
present theory with different Skyrme forces.
The smearing parameter is $\Delta =0.2$ MeV.
The dotted line are the data of ref. \cite{GSI}.}
\end{figure}

In Fig. \ref{fig1} the photo absorption cross sections for $^{120}$Sn and $^{132}$Sn are compared with
the experimental ones.
The full line represents the result of the present approach.
The dotted line is a Lorentz parametrization of the experimental data. In both cases
the shape of the calculated cross sections agrees very well with the experimental ones. 
The energies are in both cases, however, too low. 
Quantitatively, we give the same results in Table \ref{tab:t2} and compare it with the RPA results.
In Fig.\ref{fig2} the experimental cross section  of $^{132}$Sn (dotted line) is compared with the results of the RPA and the present theory where we used three different Skyrme parameterizations.
The theoretical results look similar but deviate in detail. In all three cases the energies calculated within the ETFFS approach are too low compared to the data \cite{GSI}, but the strength distribution is strongly improved compared to the RPA calculation.
In Table \ref{tab:t2} we give a more detail comparison.
The BSk5 and SLy4 parameterizations give very similar results; the widths are roughly 50\% larger compared to the one calculated in RPA and are in
good agreement with the data. The energies of the extended  theory
are more than 1.5 MeV lower than the RPA results which are already 2 MeV too low compared to the
experiment. 

\begin{figure}
{\includegraphics[width=.95\linewidth,angle=-0]{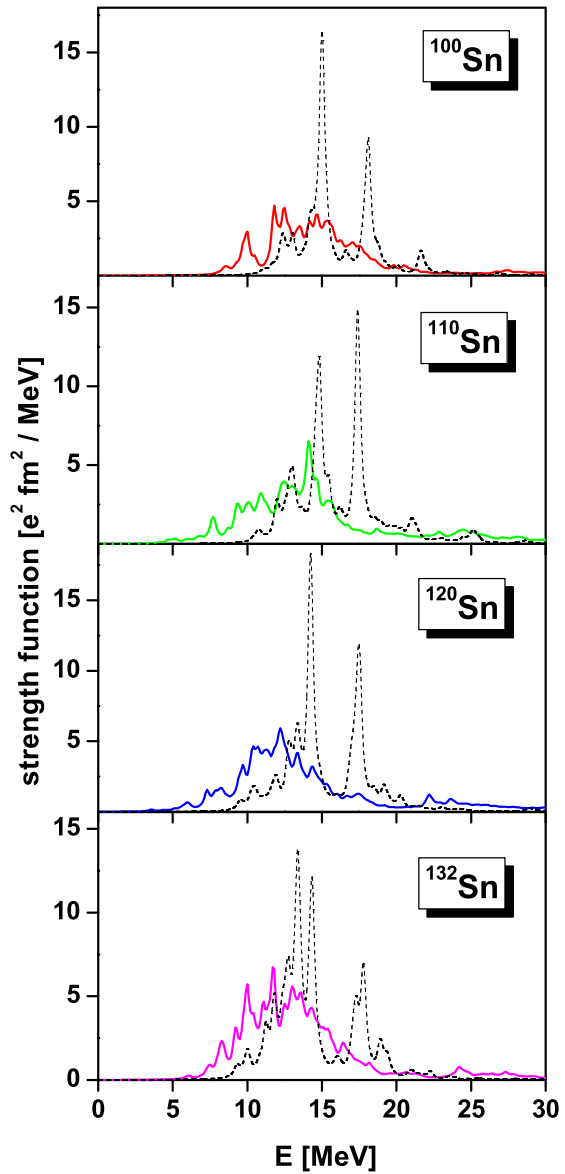}}
\caption{\label{fig3}Strength functions for GDR in $^{100}$Sn, $^{110}$Sn, $^{120}$Sn and $^{132}$Sn.
The (Q)RPA results (dashed lines) are compared with the results of the ETFFS. The SLy4 Skyrme parametrization was used.}
\end{figure}

Both effects are connected with the inclusion of the phonons. (I) The single particle strength is fragmented due to the coupling to the phonons and this effect gives rise to the spreading width which strongly improves the agreement with the experiments. (II) The phonons, however, also compress the single particle spectrum and for that reason the energies calculated in ETFFS are lower by about 1.5 MeV compared with the RPA.
The SKM* result differs from the first two parameterizations. Already on the RPA level the strength is less fragmented compared to the previous ones. If phonons are included a fraction of the narrow RPA peak survives and gives rise to a much smaller width than in the previous calculations.  The method to approximate the strength distribution by one single Lorentzian is not well justified in the RPA case
because the RPA response fragments in two clearly separated peaks. 
If one uses this procedure, one obtains the average of the two peaks and a Lorentzian width
which happens to be large . 
If the analysis includes the main peak only, the RPA width reduces to 1 MeV.
\begin{table}[!hb]
\caption{\label{tab:t2} Mean energy and width of the GDR in $^{120}Sn$ and $^{132}Sn$ calculated within
the RPA and the ETFFS. The SLy4 Skyrme parametrization was used. 
The E1 strength is fitted to a Lorentzian in the range of (8 - 22) MeV}
\begin{center}
\begin{tabular}{|c|c|c|c|c|}
\hline
  &  $E_{0}(MeV)$   & $\Gamma(MeV)$    \\
\hline
$^{120}$Sn(ETFFS) &  12.3 & 5.2\\
        (RPA) &  15.1 & 3.6\\
        (Exp) &  15.4 & 4.9\\
$^{132}$Sn(ETFFS) &  12.5 & 5.2\\ 
        (RPA) &  14.3 & 3.8\\
        (Exp) &  16.1 & 4.7\\   
\hline
\end{tabular}
\end{center}
\end{table}
In Fig. \ref{fig3} the E1 strength distributions for $^{100}$Sn, $^{110}$Sn, $^{120}$Sn and
$^{132}$Sn are shown.
In Table \ref{tab:t3} we compare (Q)RPA results  with the ones obtained with the present theory.
In all cases the energy of the present theory is shifted to lower energies by about 2 MeV,
independent of the number of neutrons, and the widths are increased by 50\%.
As the widths are increased the maximum of the cross section is decreased up to 50\%.
The strength in the two closed-shell nuclei $^{100}$Sn and $^{132}$Sn has about the same width,
despite the fact that one has added 32 neutrons.  The two closed-shell nuclei show, as expected, a compact and relatively narrow GDR compared to the two open-shell nuclei, $^{110}$Sn and $^{120}$Sn.
\begin{table}[!hb]
\caption{\label{tab:t3}Integral characteristics of the GDR in $^{132}$Sn calculated within the RPA
and with inclusion of phonons with three different Skyrme forces.
The E1 strength is fitted to a Lorentzian in the (8 - 22) MeV interval.
Data are from Ref. \cite{GSI} }
\begin{center}
\begin{tabular}{|c|c|c|c|c|}
\hline
  &  $E_{0}(MeV)$   & $\Gamma(MeV)$  \\
\hline
BSk5(ETFFS) &  12.6 & 4.4\\
BSk5(RPA)   &  14.0 & 2.9\\
SLy4(ETFFS) &  12.5 & 5.3\\
SLy4(RPA)    & 14.3 & 3.8\\
SKM*(ETFFS) &  13.1 & 3.4\\
SKM*(RPA) &  14.0 & 3.3\\
exp. &  16.1(7)& 4.7\\
\hline
\end{tabular}
\end{center}
\end{table}

We also mention that for $^{132}$Sn we solved the continuum RPA.
The nearly complete overlap with the discretized RPA shows that our discretization procedure,
together with the smearing parameter $\Delta = 200$ keV, is well justified for the present energy
range.
As the two distributions are hard to distinguish, we do not show the continuum result in
Fig. \ref{fig3}.

Our results of $^{120}$Sn and $^{132}$Sn for the GDR is qualitatively similar to the calculations by Bortignon et al.
\cite{sarchi04}. They use for their comparison the SIII parametrization which gives on the RPA level
a centroid energy  about 2 MeV higher than the SLy4 parametrization. Therefore SIII has the right tendency. 

However, the far too high incompressibility (see Table \ref{tab:t1}) was one of the motivations for the new generation of parameterizations with a different density dependence ($\alpha <1$), which we use in our calculations. Note: there is a difference between the definition of the mean energy of the total E1 distribution $E_{0}=m_{1}/m_{0}$ and the mean energy of the GDR which is is used  in Table \ref{tab:t3}. With the first definition the RPA and ETFFS results are very similar e.g. for $^{132}$Sn $ E_{0}=14.74 MeV$ (RPA) and $ E_{0}=14.47 MeV$ (ETFFS), whereas the mean energy of the GDR (Table \ref{tab:t3}) differ up to 1.8 MeV. If one compares the theoretical results with the data one has to use the latter definition because the experimentalists extract their results in the same way from their data.

\begin{figure}
{\includegraphics[width=.95\linewidth,angle=-0]{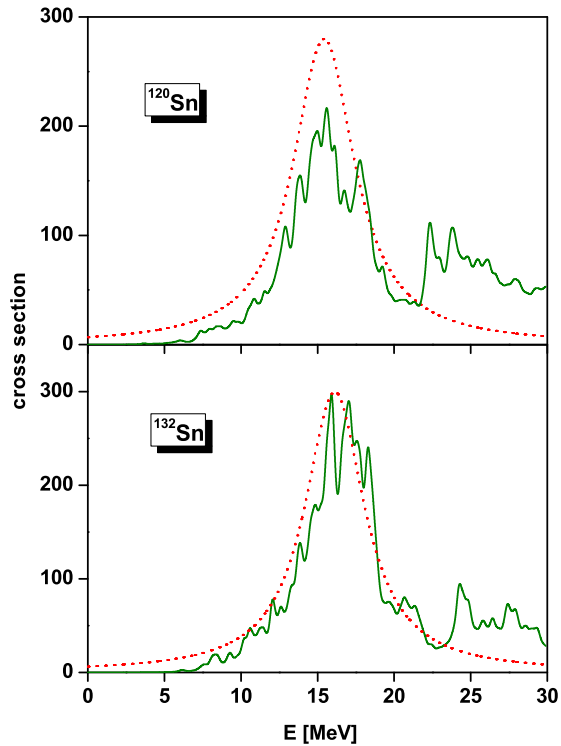}}
\caption{\label{fig4}E1 photo absorption cross section for $^{120}$Sn and $^{120}$Sn calculated
within the present theory where the isovector force has been increased by 20\% (full). 
The dotted lines indicates the lorentzian fit to the experimental data \cite{Fultz69,GSI}.}
\end{figure}

Finally we compare in Fig. \ref{fig4} the original theoretical result for the GDR in $^{120}$Sn and $^{132}$Sn with the result where the isovector force was increased by 20\%. Here we obtain good agreement with the data.  
The structures beyond 20 MeV are due to non collective ph components which will be smeared out if we include the single particle continuum.

\subsection{Collective electric isoscalar states}
Energetically low-lying collective states, such as the $3^-$ mode, are
extremely sensitive to the strength of the isoscalar particle hole interaction.
\begin{figure}
{\includegraphics[width=.95\linewidth,angle=-0]{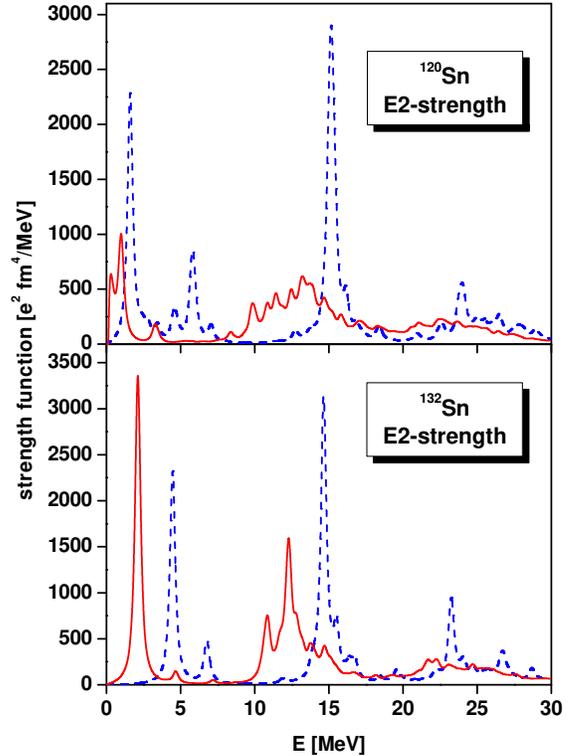}}
\caption{\label{fig5}The  strength function for $2^+$ excitation for $^{120}$Sn and $^{132}$Sn
calculated with (full) and without(dashed) Phonons.
The smearing parameter is $\Delta =0.2$ MeV.The SLy4 Skyrme parametrization was used. }
\end{figure}
In Fig. \ref{fig5} the E2 strength functions are presented for $^{120}$Sn and $^{132}$Sn.
We show the results of the RPA and the present theory.
The lowest $2^+$ in $^{120}$Sn is below zero and, as well, the isoscalar giant electric quadrupole
resonance is too low compared with the data.
In $^{132}$Sn the isoscalar electric quadrupole resonance has not yet been measured.
In Table \ref{tab:t4} the theoretical energies of the low lying collective $2^+$ and $3^-$ states
are compared with the experiments.
 The RPA results for the $2^+$ modes  are in nice agreement with the data for both isotopes.
The $3^-$mode in $^{132}$Sn is underestimated by 1.35 MeV.
In the presence of phonons, the collective $3^-$ modes are pulled down by approximately 1.5 MeV.
The $2^+$ mode in $^{132}$Sn is shifted downwards by 2.36 MeV, and the $2^+$ mode in $^{120}$Sn
is even shifted below the ground state energy.
Here, however, one has to bear in mind that in $^{120}$Sn the magnitude of the pairing gaps plays
an important role.

The reduction of the excitation energies of the collective isoscalar modes with respect to the
random phase approximation has two sources, the modified level density in the vicinity of the
Fermi level, and the strength of the isoscalar particle hole interaction.
The compression of the level density due to phonons can be understood qualitatively by
second order perturbation theory: The intermediate states generated by the phonon degree of freedom
have larger excitation energies than the single particle level obtained in the mean field
approximation. Therefore phonons introduce an additional attractive interaction, whatever the details
of the phonon-particle coupling are, which increases the level density near the Fermi level.
The only way to compensate the reduced level spacing between particle and hole states is to
reduce the strength of the isoscalar particle hole interaction in comparison with the
interaction used in a random phase calculation.

Similar arguments hold for the collective isovector modes discussed above. The repulsive
isovector parts of the particle hole interaction have to be increased compared with
the strength used in the random phase approximation in order to compensate the 
phonon effects. 
This conclusion is independent on details of the theoretical framework.
\begin{table}[!hb]
\caption{\label{tab:t4}Low-lying collective $2^+$ and $3^-$ states
in $^{120}$Sn and $^{132}$Sn calculated
within and without phonons.The SLy4 Skyrme parametrization was used. }
\begin{center}
\begin{tabular}[b]{|c|c|c|c|c|}
\hline
  & \multicolumn{2}{c|}{$^{120}$Sn}
  & \multicolumn{2}{c|}{$^{132}$Sn} \\
\hline
  & 2$^+$ [MeV] & 3$^-$ [MeV]  & 2$^+$ [MeV] & 3$^-$ [MeV]  \\
\hline
without phonons & 1.60 & 2.60 & 4.46  & 3.00 \\
with phonons & --- & 1.3 & 2.10  & 1.50 \\
exp. & 1.17 & 2.40 & 4.04 & 4.35 \\
\hline
\end{tabular}
\end{center}
\end{table}

\section{Conclusion}
In the present paper we report self-consistent nuclear structure calculations
performed in the Extended Theory of Finite Fermi Systems. 
Skyrme parameterizations were used, which reproduce not only ground state properties of nuclei but also excited collective states within the RPA.
In comparison with RPA calculations, the widths of the giant resonances are brought into better
agreement with the experimental data. The centroid energies of the resonances are shifted, however.
This finding implies the necessity of a renormalization of Skyrme interactions for applications beyond the
mean field level. Our investigation suggests two major renormalizations: a reduction of the strength of the
isoscalar particle-hole interaction and an increase of the isovector strength.

\section{Acknowledgment}
We thank Victor Tselyaev for many discussion concerning the QTBA.
We also thank Sergey Tolokonnikov for useful comments on the energy-density functional.
One of us (JS) thanks Stanislaw Dro\.zd\.z for many discussions and the Foundation for Polish
Science for financial support through the
\emph{Alexander von Humboldt Honorary Research Fellowship}.
The work was partly supported by the DFG and RFBR grants Nos.GZ:432RUS113/806/0-1 and 05-02-04005
and by the INTAS grand No.03-54-6545.


\begin{references}
\bibitem{Thielemann:2001rn}F.K. Thielemann {\it et al.}, Prog. Part. Nucl. Phys. 46 (2001) 5.
\bibitem{Stan}S. Dro\.zd\.z, V. Klemt,J. Speth and W. Wambach, Phys. Rep. 197 (1990) 1.
\bibitem{Tsel89}V.I. Tselyaev, Sov. J. Nucl. Phys. 50 (1989) 780.
\bibitem{Kam91}S.P. Kamerdzhiev and V.I. Tselyaev, Bull. Acad. Sci. USSR, Phys. Ser. 55 (1991) 45.
\bibitem{kst93}S.P. Kamerdzhiev, J. Speth, G. Tertychny and V. Tselyaev, Nucl. Phys. A 555 (1993) 90.
\bibitem{kstw93}S.P. Kamerdzhiev, J. Speth, G. Tertychny and J. Wambach, Z. Physik A 336 (1993) 253.
\bibitem{Vic}V.I. Tselyaev, arXiv:nucl-th/0505031; Phys.Rev.C (to be published);
             E.V. Litvinova and V.I. Tselyaev, arXiv:nucl-th/0512030.
\bibitem{rev04}S. Kamerdzhiev, J. Speth and G. Tertychny, Phys. Rep. 393 (2004) 1.
\bibitem{gruemmer}F.Gr\"ummer and J.Speth, J.Phys.G: Nucl. Part. Phys. 32 (2006) R193.
\bibitem{Bort01}G. Colo, P.F. Bortignon,  Nucl. Phys. A 696 (2001) 427.
\bibitem{BB59}G.E.Brown and M. Bolsterli, Phys. Rev. Lett. 3(1959) 472.
\bibitem{Ring74}P. Ring and J. Speth, Nucl. Phys. A 235 (1974) 315.
\bibitem{Zawischa}D. Zawischa, J. Speth and D. Pal, Nucl. Phys. A 311 (1978) 445.
\bibitem{Speth07}V. Tselyaev, J. Speth, F. Gr\"ummer, S. Krewald, A. Avdeenkov, E. Litvinova and G. Tertychny,
                 Phys. Rev. C 75 (2007) 014315.
\bibitem{Epelbaum:2005pn}E. Epelbaum, Prog. Part. Nucl. Phys. 57 (2006) 654
\bibitem{Bender:2003jk}M. Bender, P.H. Heenen and P.G. Reinhard, Rev. Mod. Phys. 75 (2003) 121.
\bibitem{Bertsch:2004us}G.F. Bertsch, B. Sabbey and M. Uusnakki, Phys. Rev. C 71 (2005) 054311.
\bibitem{MAHAUX}C. Mahaux, P.F. Bortignon, R.A. Broglia and C.H. Dasso, Phys. Rep. 120 (1985) 1.
\bibitem{SLy4}E. Chabanat, P. Bonche, P. Haensel, J. Meyer, R. Schaeffer, Nucl. Phys. A 635 (1998) 231. 
\bibitem{terasaki1}J. Terasaki, J. Engel, M. Bender et al., Phys. Rev. C 71 (2005) 034310.
\bibitem{goriely1}S. Goriely, E. Khan, M. Samyn, Nucl. Phys. A 739 (2004) 331.
\bibitem{terasaki2}J. Terasaki, Phys. Rev. C 74 (2006) 044301.
\bibitem{Bender:2005ri}M. Bender, G.F. Bertsch and P.H. Heenen, Phys. Rev. C 73 (2006) 034322.
\bibitem{LR06}E. Litvinova, P. Ring and D.Vretenar, preprint (January 2007).
\bibitem{LRT06}E. Litvinova, P. Ring and V. Tselyaev, preprint (September 2006).
\bibitem{bennaceur}K. Bennaceur and J. Dobaczewski, Comp. Phys. Comm. 168 (2005) 96. 
\bibitem{speth}S.O. B\"ackman, A.D. Jackson, J. Speth, Phys.Lett. B 56 (1975) 209.
\bibitem{KKSF77}S. Krewald, V. Klemt, J. Speth and A. Faessler, Nucl. Phys. A 281 (1977) 166.
\bibitem{BSk5}M. Samyn, S. Goriely, P.-H. Heenen, J.M. Pearson and F. Tondeur, Nucl. Phys. A 700 (2002) 142.
\bibitem{SkM*}J. Bartel, P. Quentin, M. Brack, C. Guet and H.-B. H{\aa}kansson, Nucl. Phys. A 386 (1982) 79.
\bibitem{tselyaev00}V.I. Tselyaev, Bull. Rus. Acad. Sci. Phys. 64 (2000) 434.
\bibitem{Fultz69}S.C. Fultz, B.L. Berman, J.T. Caldwell, R.L. Bramblett and M.A. Kelly, Phys. Rev. 186 (1969) 1255.
\bibitem{GSI}P. Adrich et al., Phys. Rev. Lett. 95 (2005) 132501-1.
\bibitem{PR06}E. Litvinova and P. Ring, Phys. Rev. C 73 (2006) 044328.
\bibitem{sarchi04}D.Sarchi,P.F.Bortignon and G.Colo, Phys.Lett. B601 (2004) 27.
\end{references}
\end{document}